\documentclass[
 aps, pra,
 amsmath,amssymb,
 11pt,
 final,
tightenlines,
 twoside,
 onecolumn,
 nofloats,
nofootinbib,
 superscriptaddress,
showkeys,
showkeywords,
 ]
{revtex4-2}

\usepackage[T2A]{fontenc}
\usepackage[utf8]{inputenc}
\usepackage[russian,english]{babel}
\usepackage{graphicx}
\usepackage{dcolumn}
\usepackage{bm}

\usepackage{ulem}

\input{maik.rty}

\setcitestyle{authoryear,round}
\def\squareforqed{\hbox{\rlap{$\sqcap$}$\sqcup$}}

\def\sq{\ifmmode\squareforqed\else{\unskip\nobreak\hfil
\penalty50\hskip1em\null\nobreak\hfil\squareforqed
\parfillskip=0pt\finalhyphendemerits=0\endgraf}\fi}

\def\utw{\smash{\rlap{\lower5pt\hbox{$\sim$}}}}

\def\udtw{\smash{\rlap{\lower6pt\hbox{$\approx$}}}}

\def\diameter{{\ifmmode\mathchoice
{\ooalign{\hfil\hbox{$\displaystyle/$}\hfil\crcr
{\hbox{$\displaystyle\mathchar"20D$}}}}
{\ooalign{\hfil\hbox{$\textstyle/$}\hfil\crcr
{\hbox{$\textstyle\mathchar"20D$}}}}
{\ooalign{\hfil\hbox{$\scriptstyle/$}\hfil\crcr
{\hbox{$\scriptstyle\mathchar"20D$}}}}
{\ooalign{\hfil\hbox{$\scriptscriptstyle/$}\hfil\crcr
{\hbox{$\scriptscriptstyle\mathchar"20D$}}}}
\else{\ooalign{\hfil/\hfil\crcr\mathhexbox20D}}%
\fi}}

\def\be{\begin{equation}}
\def\ee{\end{equation}}
\def\ba{\begin{eqnarray}}
\def\ea{\end{eqnarray}}

\def\msun{M_\odot}

\def\ltsima{$\; \buildrel < \over \sim \;$}
\def\simlt{\lower.5ex\hbox{\ltsima}}
\def\gtsima{$\; \buildrel > \over \sim \;$}
\def\simgt{\lower.5ex\hbox{\gtsima}}

\def\zsun{Z_\odot}

\usepackage[dvipsnames]{color}

\begin{document}

\selectlanguage{english}


\title{Infrared and X-ray emission of a supernova remnant in a clumpy medium\footnote{published in Astronomy Reports, 69, 1 (2025), doi:  10.1134/S1063772925701495}}

\author{\firstname{S.~Yu.}~\surname{Dedikov}}
 \email{s.dedikov@asc.rssi.ru}
 \affiliation{Lebedev Physical Institute of Russian Academy of Sciences, 53 Leninskiy Ave., 119991, Moscow, Russia}

\author{\firstname{E.~O.}~\surname{Vasiliev}}
 \email{eugstar@mail.ru}
 \affiliation{Lebedev Physical Institute of Russian Academy of Sciences, 53 Leninskiy Ave., 119991, Moscow, Russia}

\begin{abstract}
The infrared (IR) to X-ray luminosity ratio (IRX) is an indicator of the role of the dust plays in cooling of hot gas in supernova remnants (SNR). Using the 3D dynamics of gas and interstellar polydisperse dust grains we analyze the evolution of SNR in the inhomogeneous medium. We obtain spatial distributions of the surface brigthness both of the X-ray emission from hot gas inside SNR and the IR emission from the SNR swept-up shell, as well as, the average gas temperature in the SNR, $T_X$. We find that the IRX changes significantly (by a factor of $\sim 3-30$) as a function of impact distance within the SNR and its age. In a low inhomogeneous medium the IRX drops rapidly during the SNR evolution.  On the other hand, if large inhomogeneities are present in the medium, the IRX is maintained at higher levels during the late SNR evolution at radiative phase due to replenishment of dust in the hot gas by incompletely destroyed fragments behind the shock front. We show that the onset of the radiative phase determines the evolution of the $T_X - {\rm IRX}$ diagram. We illustrate that decreasing gas metallicity or density leads to high values of temperature and IRX ratio. We discuss how our results can be applied to the observational data to analyse the SNR older than 10~kyr (i.e. when the mass of the swept-up dust in the shell is expected to exceed that produced in the SNR) in the Galaxy and Large Magellanic Cloud.
\\
\\
{\bf Keywords}: {galaxies: ISM -- ISM: shells -- shockwaves -- supernova remnants -- dust}
\\
\\
\end{abstract}

\maketitle

\section{INTRODUCTION}

\noindent
During adiabatic expansion of a supernova remnant (SNR), the shock front moves through the interstellar medium at a speed of over 200~km/s, so the gas temperature in the swept-up shell reaches several million degrees or more. Dust particles in the medium fall into the hot gas, where they are capable of both significantly cooling the gas \citep{Ostriker1973,Burke1974,Silk1974,Smith1996} and rapidly destroying due to thermal sputtering \citep{Draine1979b}. Therefore, SNRs are a laboratory for studying the efficiency of these processes.

The contribution of dust to the cooling of hot dusty plasma is determined from the ratio of energy losses in
infrared (IR) emission of dust particles $L_{IR}$ to radiation in the X-ray range from atomic processes $F_X$, i.e., due to bremsstrahlung and emission in the lines of highly ionized metal ions \citep{Dwek1987a,Dwek1987b}:
\be
 {\rm IRX} \equiv L_{IR}/L_X. 
\label{irx}
\ee

The IRX ratio was found to be significantly greater than one for several SNRs \citep{Dwek1987b,Graham1987}, which is in favor of the dominant role of dust in cooling the hot gas. Later, \citet{Seok2013,Seok2015} used Spitzer and AKARI IR observations \citep{Meixner2006,Onaka2007} and Chandra X-ray data\footnote{http://hea-www.cfa.harvard.edu/ChandraSNR/} to obtain IRX values for a larger number of remnants in the Large Magellanic Cloud (LMC) and compared them with remnants in the Galaxy. They concluded that cooling on dust is more efficient, although cooling due to atomic processes is not negligible. The IRX ratio in the LMC remnants was found to be systematically lower than in the Galactic remnants. Apparently, this reflects the properties of the interstellar medium in LMC, in particular, the lower dust content in the gas. A simple comparison of the observed values with the theoretical dust cooling function \citep{Dwek1987a,Dwek1992} showed a difference of more than an order of magnitude. This can be explained by the dust destruction or local variations of dust in the medium ahead of the shock front. The spatial distribution of gas and dust affects the morphology of the SNR and, consequently, the IRX ratio, as indicated by \citep{Koo2016} during the study of 20 SNRs in the Galaxy. \citet{Dwek1987b} noted the possible influence of the medium properties and the interaction of the SNR with the surrounding clouds on the IRX value.

Note that the analysis mainly included fairly young SNRs with an age from several hundred to thousands of years \citep[see, e.g.][]{Seok2015}. By this time, the SNR shell does not accumulate too much mass, and the hot gas of the ejecta contains dust particles produced at the early phases of the remnant’s evolution \citep[see, e.g.][]{Todini2001}. This dust is not completely destroyed in the inner regions of the SNRs \citep[see, e.g.][]{Micelotta2016,Micelotta2018,Slavin2020,Kirchschlager2019,Priestley2021b}, and its radiation can contribute to the IR luminosity. At later times, the shell already contains a significantly larger mass of swept-up interstellar dust and its radiation will probably dominate the IR luminosity of the remnant.

In this paper, we consider the evolution of radiation in such late SNRs, and investigate the IRX ratio as the remnant expands in an inhomogeneous medium. In Section 2, we describe the model and initial conditions. In Section 3, we present the results. In Section 4, we discuss the application of the results and their consequences. In Section 5, we briefly summarize the main results.

\section{DESCRIPTION OF THE MODEL}

\noindent
Let us consider the relationship between the emission properties of hot X-ray gas and interstellar polydisperse dust particles emitting in the IR range in the SNR expanding in a non-uniform (clumpy) medium. To do this, we self-consistently take into account the cooling of the gas due to dust emission in the equations of gas dynamics and dust particle transport \citep{vs2024}. In what follows, we give a brief description of our model of gas
and dust dynamics in the SNR \citep[see for more details][]{Dedikov2024}. 

\subsection{Dynamics of Gas and Dust}

\noindent
To obtain a non-uniform gas density field, we use the pyFC module\footnote{https://bitbucket.org/pandante/pyfc/src/master/} \citep{Lewis2002}, which allows generating “fractal cubes” with a lognormal amplitude distribution and a Kolmogorov spatial spectrum with a power-law index $\beta$ = 5/3. The density field is characterized by the mean value $\langle n\rangle$, the standard deviation of the log-density $\sigma$, and the wave number $k_{min}$, which determines the maximum spatial size of fluctuations. 

The mean gas density is set to be $\langle n\rangle = 1$~cm$^{-3}$, the variance $\sigma$ is equal to 0.2 for a low non-uniform medium and 2.2 for a strongly non-uniform medium. With such parameters, the density contrast
for 2$\sigma$ fluctuations reaches a factor $\simgt 1.5$ for a smaller $\sigma$ value and $\simgt 80$ for a high $\sigma$. In the case of a strongly non-uniform medium with $\langle n\rangle = 1$~cm$^{-3}$ the density in some fragments exceeds 100~cm$^{-3}$. The maximum size of density fluctuations in the models under consideration is 6~pc, which corresponds to $k_{min}=16$ for a grid ($96$~pc)$^3$ and a number of cells equal to $256$ in each spatial direction. A cell size of 0.375~pc is sufficient for an adequate consideration of the dynamics of the SNR in an inhomogeneous medium \citep[see the discussion in][]{Dedikov2024}.

At the initial time, the gas is in equilibrium ($\rho T$ = const) with zero velocities of gas and dust particles in the surrounding interstellar gas. The dust-to-gas density ratio is taken to be ${\cal D} = 0.01$ for the solar metallicity and is assumed to be proportional to the metallicity value. Initially, the size distribution of interstellar dust particles followed a power law with a slope $-3.5$ \citep{Mathis1977} in the range of $30-3000$~\AA, divided into 11 equal bins on a logarithmic scale. The minimum dust particle size in the calculations is 10~\AA. 

To take into account radiation losses, the calculations use a nonequilibrium cooling function \citep{v11,v13} obtained for an isochoric process of gas cooling from от $10^8$~K to 10~K, including the ionization kinetics of
all ionic states of the following chemical elements: H, He, C, N, O, Ne, Mg, Si, and Fe. The gas heating is
set to stabilize the medium with a pressure $nT \simeq 10^4$cm$^{-3}$K, not perturbed by the shock wave from
the SNR.

When the supernova explodes, mass and energy are injected into a small region. The size of this region is
1.5~pc. The energy of the supernova is $10^{51}$~erg and is added as thermal energy. The mass of the injected gas
and metals is 30 and 10~$\msun$, respectively \citep[$M_Z \sim 0.3M_{SN}$ for massive supernovae, see][]{Woosley1995}. The evolution of dust particles produced in the early phases of the SNR expansion will be considered in a separate paper.

For the numerical solution of the gas dynamics equations, the software package \citep{vns2015,vsn2017} is used, based on  the unsplit total variation diminishing (TVD) approach that provides high-resolution capturing of shocks and prevents unphysical oscillations, and the Monotonic Upstream-Centered Scheme for Conservation Laws (MUSCL)-Hancock scheme with the Haarten-Lax-van Leer-Contact (HLLC) method \citep[see e.g.][]{Toro2009} as approximate Riemann solver.

To describe the dust dynamics, this package has been upgraded \citep[see Appendix A in][]{vs2024}, using the “superparticle” method proposed in \citep{Youdin2007,Mignone2019,Moseley2023}. For each superparticle, the equations of motion are solved taking into account the mutual influence on the gas due to friction forces \citep{Epstein1924,Baines1965,Draine1979a} and the equation for the evolution of the radius of a particle due to thermal and kinetic sputtering \citep{Draine1979b}. A superparticle is a conglomerate of identical microparticles -- dust particles (grains). To track dust transfer in the medium, at least one superparticle is placed in each numerical cell for each bin of the particle size distribution. Thus, the total number of superparticles for the adopted grid will be $256^3 N_{s} \sim 184$~millions, where $N_{s}$ is the number of bins in the grain size distribution.

Note that thermal sputtering dominates in hot gas, but at $T\simlt 10^5$~K the destruction of dust due to collisions with each other (shattering) can become significant \citep{Jones1996,Hirashita2009,Murga2019}. The characteristic time of this process varies from several hundred thousand to tens of millions of years \citep[e.g.,][]{Hirashita2009,Martinez2019,Kirchschlager2022}. The presence of magnetic fields can enhance the influence of this process \citep[see, e.g.,][and references therein]{McKee1987,Seab1987}. Although recent studies show a rather complex behavior of particles in a magnetic field and indicate the possible protection of grain destruction during collisions with each other \citep{Moseley2023}. 

\begin{figure*}
\center
\includegraphics[width=8cm]{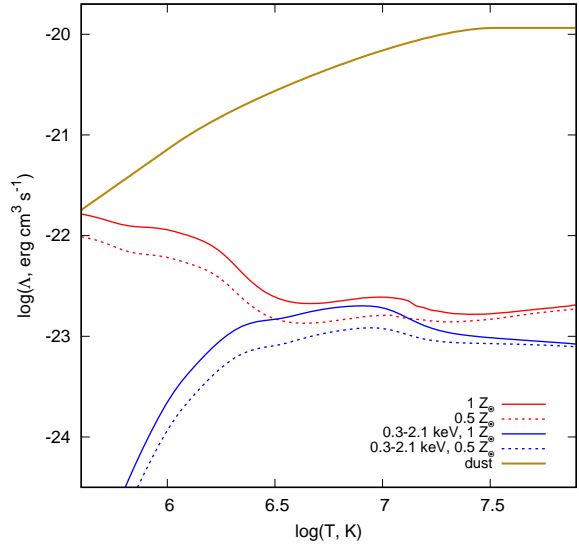}
\caption{
Gas cooling rates for metallicity $1\zsun$ (red solid line) and $0.5\zsun$ (red dashed line) taken from \citet{v13} and losses due to X-ray radiation in the range of $0.3-2.1$~keV (blue lines). The yellow line shows the losses due to IR radiation of dust particles with a power-law size distribution with a slope $-3.5$ \citep{Mathis1977} in the range of $30-3000$~\AA.
}
\label{fig-cool}
\end{figure*}

\subsection{ Cooling on Dust}

\noindent
Cooling due to IR radiation from dust particles is added to the equation for gas energy:
\be
 {\partial E_{gas} \over \partial t} + ... = ... - L_{d,IR}. 
\ee
Dust particles in the hot gas of the SNR are heated mainly by collisions with electrons \citep{Draine1979a} and emit this energy in the IR range. At equilibrium, the rates of cooling and heating are equal: $L_{d,IR}(a, T_{eq}) = H_{coll}(a,T_{g} ,n_{e}) n_d(a)$, where $a$ is the particle size, $T_g$ is the gas temperature, $n_e$ is the electron number density, $n_d(a)$ is the number density of dust particles of size $a$, \citep[e.g.,][]{Dwek1992}. The heating rate of a dust particle of radius $a$ is \citep{Dwek1992}: 
\be
 H(a,T_g,n_e) = 5.38 \times 10^{-18} n_e a_{\mu{\rm m}} T_g^{3/2} h(a,T_g)
\ee
where $h(a,T_g) = 1$ for $T_g<T_1(a)$ and $h(a,T_g) = [T_g/T_1(a)]^{-3/2}$ for $T_g>T_1(a)$, $T_1(a) = 3\times 10^5 (a/0.0005\mu{\rm m})^{3/4}$. Figure~\ref{fig-cool} presents the cooling function for dust particles with a size distribution according to a power law with a slope $-3.5$ \citep{Mathis1977} in the range of $30-3000$~\AA. The functions of gas cooling due to atomic processes in hot plasma \citep{v13} and losses due to X-ray radiation in the range of $0.3-2.1$~keV are shown as well.

Note that small particles with a size of $a \sim 30$~\AA \ can experience strong temperature fluctuations in a hot gas \citep[e.g.,][]{Dwek1986}. Taking this mechanism into account requires constructing the temperature distribution functions of dust particles \citep[e.g,][]{Drozdov2019} based on direct modeling of gas-dust collisions using the Monte Carlo method, which is difficult to do self-consistently in the three-dimensional joint dynamics of dust and gas. The calculation of the IR emission from small grains taking into account stochastic heating in the three-dimensional dynamics of the SNR has showed that their fraction in the total IR luminosity can be remarkable only in the first 10~kyr after SN explosion \citep{Drozdov2025}. The contribution to the IR luminosity from polycyclic aromatic hydrocarbons (PAHs) is not taken into account, since it does not exceed several percent for the typical values of the PAH content of $q_{PAH} \sim 1$\% and the flux of external ultraviolet radiation in the local interstellar medium $U\sim 1$ \citep{Draine2007}.

\section{RESULTS}

\noindent
We start our consideration of the emission properties of gas and dust in the SNR to an age of 10~kyr, which is associated with a possible remarkable contribution from dust produced (injected) by the supernova, the dynamics of which are not taken into account in our calculations. By this age, the mass of the accumulated interstellar dust turns out to be several times higher \citep[e.g.,][]{Slavin2020,vs2024,Dedikov2024}. We complete our calculations by the age of 100~kyr, when the gas in the SNR, expanding in a medium with a density of $\sim 1$~cm$^{-3}$, is cooled effectively and the mass of hot gas in the SNR, emitting in the X-ray range, becomes insignificant.

\subsection{Evolution of the Remnant}
\label{sec-evol}

\noindent
During expansion in an inhomogeneous medium, the SN shell interacts with gas of different densities, and the shock wave penetrates into regions of lower density at a higher speed and, conversely, slows down in dense clouds \citep{Korolev2015,Slavin2017,Wang2018}. In a medium with a low level of perturbations, the shape of the SNR is close to spherical. With increasing dispersion of density perturbations, the outer surface of the shell becomes highly jagged, and numerous fragments are preserved behind the front. The features of the evolution of the SNR are described in detail earlier \citep{Dedikov2024}. 

Due to their inertia, interstellar grains penetrate far beyond the shock front and enter the gas with
$T\simgt 10^6$K и $n\simlt 0.1$~cm$^{-3}$ \citep{vs2024,Dedikov2024}. Under these conditions, the particles are subjected to effective collisions with protons and other nuclei, which lead to losses of mass and energy. As the SNR expands, the gas cools. The shell slows down and dust particles located far behind the shock front can catch up with it and pass from the hot gas of the ejecta into the cold gas of the shell \citep{vs2024}. In this case, their emissivity is significantly reduced.

The dust mass in the shell increases as the SNR expands \citep{Dedikov2024}. Its contribution to cooling dominates only at $T\simgt 10^7$~K in gas with solar metallicity ($L \sim {\cal D} L_d$, Figure.~\ref{fig-cool}). Such temperatures in the SNR are reached during the first 10--20~kyr and later in the innermost highly rarefied parts of the remnant. In gas with $T\sim 10^6-10^7$K, dust cooling remains remarkable but not dominant. Thus, taking into account dust cooling leads to a decrease in the remnant radius compared to the case of cooling only due to atomic processes in the hot plasma. The difference in the shell size gradually increases. By the age of 100~kyr (the end of the calculation), the radius of the remnant expanding in a homogeneous medium decreases by $\sim 3$~pc or about 10\% of the current radius. With increasing inhomogeneity, the difference in the values of the average radius decreases.

\begin{figure*}
\center
\includegraphics[width=8cm]{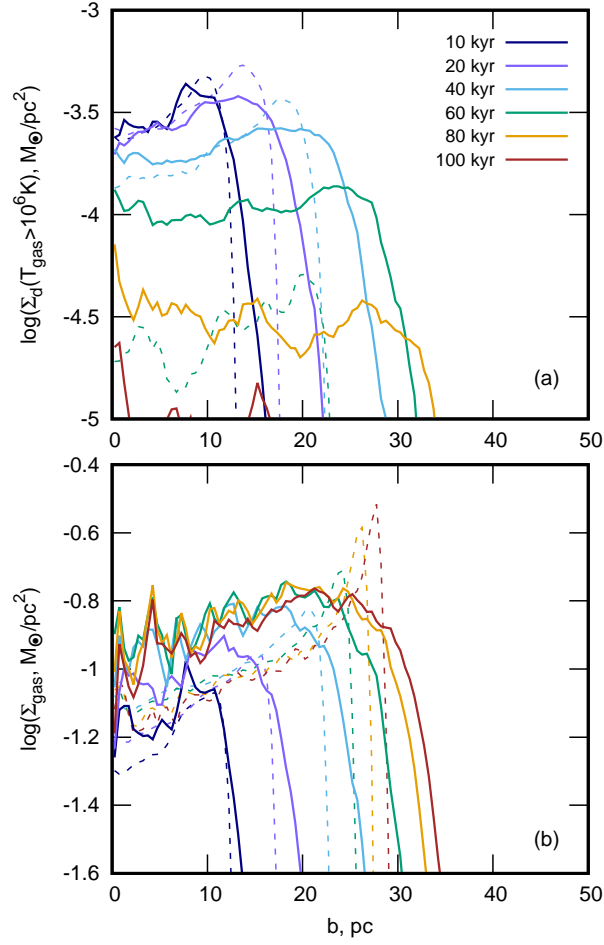}
\caption{
Surface densities of dust in hot gas with $T_{gas}>10^6$K (a) and gas (b) from the impact distance ($b=0$ corresponds to the location of the supernova explosion). The distributions for the SNR in an inhomogeneous medium with low dispersion ($\sigma=0.2$) are shown by dashed lines, with high density dispersion ($\sigma=2.2$) by solid lines. The color of the line corresponds to the age of the remnant.
}
\label{fig-surf-den}
\end{figure*}

\begin{figure*}
\center
\includegraphics[width=8cm]{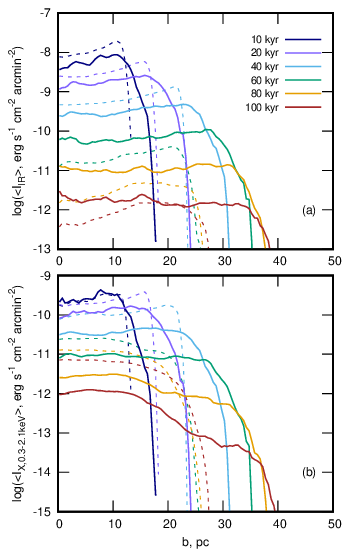}
\caption{
Surface brightnesses of IR emission from dust (a) and X-ray emission from gas in the range of $0.3-2.1$~keV (b) depending on the impact distance $b$. The distributions for the SNR in an inhomogeneous medium with low dispersion ($\sigma=0.2$) are shown by dashed lines, with high density dispersion ($\sigma=2.2$) by solid lines. The line color corresponds to the age of the remnant.
}
\label{fig-surf-b}
\end{figure*}

\subsection{Distribution of Gas and Dust} 
\label{sec-distr}

\noindent
Let us consider the 2D distribution of the values $A(y,z)$, obtained by summing up along the line of sight in the $x$ direction, averaged over the annular layers with the impact radius $b$:
\be
 \langle A(b) \rangle = {{\sum A(y,z) \Delta S(b)} \over {\sum \Delta S(b) }}
\label{eq-ave}
\ee
where $b^2 = (y-y_0)^2 + (z-z_0)^2$, $(y_0,z_0)\equiv(0,0)$ are the coordinates of the supernova explosion origin, $\Delta S(b)$ is the area of the ring. 

Figure~\ref{fig-surf-den} show the radial profiles of the surface density of dust (a) and gas (b) from the impact distance ($b=0$ corresponds to the SN explosion origin). In a low inhomogeneous medium, the SN shell is clearly defined. At the periphery, the line-of-sight intersects the shell tangentially. Therefore, the surface density of gas $\Sigma_{gas}$ (dashed lines) increases at large distances from the center of the SNR $b$ and reaches a maximum approximately at the distance equal to the size of the remnant. After the onset of the radiation phase ($t\sim 40$~kyr), the shell becomes thinner and denser. The maximum value of $\Sigma_{gas}$ increases for longer age. As noted above, in a strongly inhomogeneous medium, the shock front propagates between dense fragments, bending around and partially destroying them, so a significant part of the gas mass contained inside them remains far behind the shock wave. The line of sight, even with a small $b$, intersects them and the surface density profile of the gas (solid lines) turns out to be flatter compared to the evolution in a low inhomogeneous medium.

Dust particles located in a hot gas with $T_{gas}>10^6$K are effectively heated and emit in the IR range
(Figure~\ref{fig-surf-b}a). The surface density profiles of this dust break off at noticeably smaller impact radii (Figure~\ref{fig-surf-den}a) than the gas density profiles (panel b), because, firstly, the particles are destroyed in such aggressive conditions, and secondly, they pass into a gas with a lower temperature after the onset of the radiation phase, since the supernova shell slows down and particles retaining a higher velocity overcome the dense and cold shell \citep{vs2024,Dedikov2024}. In a strongly inhomogeneous medium, dust coming from weakly destroyed fragments located in a hot gas partially compensates for the losses, and in this case the dust density decreases more slowly (cf. solid and dashed lines in Figure~\ref{fig-surf-den}a).

Gas with $T>10^6$K emits a significant part of its energy in the X-ray range (Figure~\ref{fig-surf-b}b). During evolution in a low inhomogeneous medium before the onset of the radiation phase (the age of the remnant is less than 40~kyr), such hot gas fills the entire remnant -- the ejecta and the shell. Then the shell cools rapidly and it remains only in the ejecta, where the temperature drops slower. Therefore, even for an age of 100~kyr, the surface brightness in the X-ray range remains an order of magnitude lower than it was in the
adiabatic phase at $\sim 10-30$~kyr.

During expansion in a strongly inhomogeneous medium in a young ($\sim 10$~kyr) SNR, the shock wave completely destroys clouds in the environment and the surface brightness has the same level as in the case of small inhomogeneities. Later, the shock wave flows around a part of the clouds without destroying them completely, the hottest gas remains in the central region of the remnant, it is trapped by fragments of destroyed clouds \citep{Korolev2015}. Therefore, the surface brightness of a remnant evolving in a strongly inhomogeneous medium decreases more rapidly with age, and at large impact distances this drop increases.

\begin{figure*}
\center
\includegraphics[width=12.5cm]{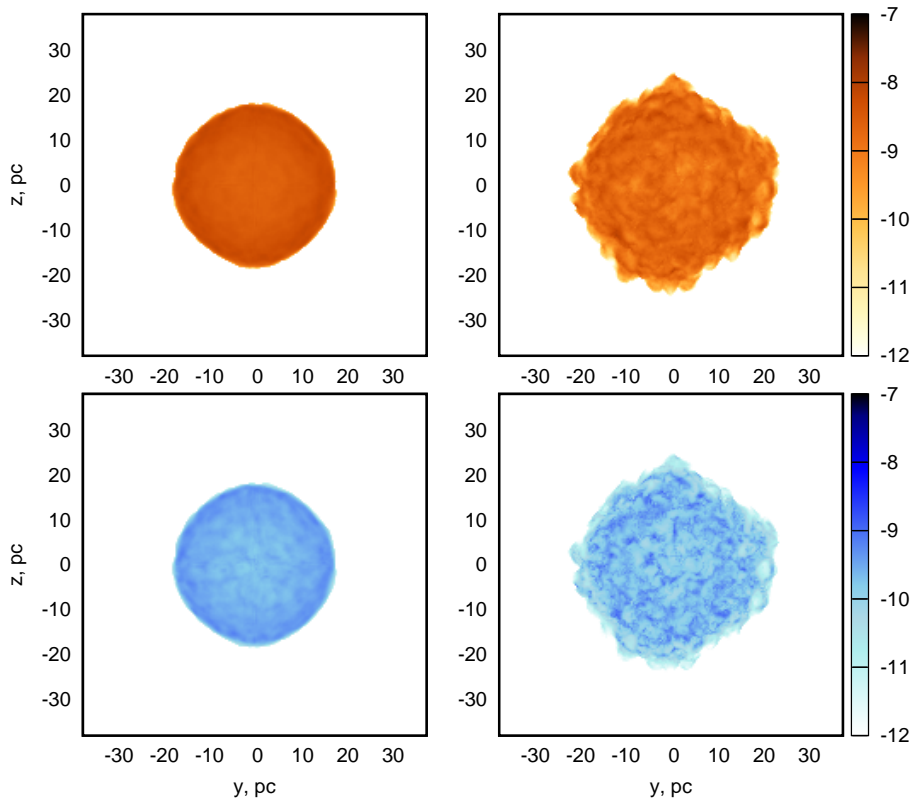}
\caption{
Surface brightness maps of IR emission from dust (upper row) and X-ray emission from gas in the range of $0.3-2.1$~keV (lower row) for a 20 kyr SNR in an inhomogeneous medium with density dispersion $\sigma=0.2$ (left column) and $\sigma=2.2$ (right column). The color scale shows the logarithm of the brightness in units of [erg~s$^{-1}$~cm$^{-2}$~arcmin$^{-2}$].
}
\label{fig-surf-b-maps}
\end{figure*}

\subsection{Surface Brightness} 
\label{sec-sb}

\noindent
Figure~\ref{fig-surf-b} shows the surface brightness profiles of dust (panel a) and hot gas (panel b). The IR emission distributions of dust weakly depend on the impact distance, especially during the expansion of the remnant in a strongly inhomogeneous medium (solid lines). Some increase in brightness toward the periphery of the remnant during evolution in a nearly homogeneous gas is due to a significant concentration of interstellar dust particles in the shell. Before the onset of the radiative phase (age less than $\sim 40$~kyr), similar behavior is observed for the surface brightness profiles of the hot gas emission in the soft X-ray range (panel b). For example, Figure~\ref{fig-surf-b-maps} presents the surface brightness maps in the IR and X-ray emission from the SNR.

The differences become apparent later and are due to faster cooling of the gas in regions close to the SN shell. In the inner part of the remnant, the gas also cools down and, although its density is lower, its emissivity still remains sufficiently high due to the slow cooling of the gas in rarefied regions. This can be seen most clearly for the age of $\simgt 80$~kyr in the model for a strongly inhomogeneous medium. Note that the effect of “locking” hot gas by half-destroyed clouds in the central region of the remnant during evolution in environments with high mass loading also plays a role here \citep{Korolev2015}. 

\begin{figure*}
\center
\includegraphics[width=8cm]{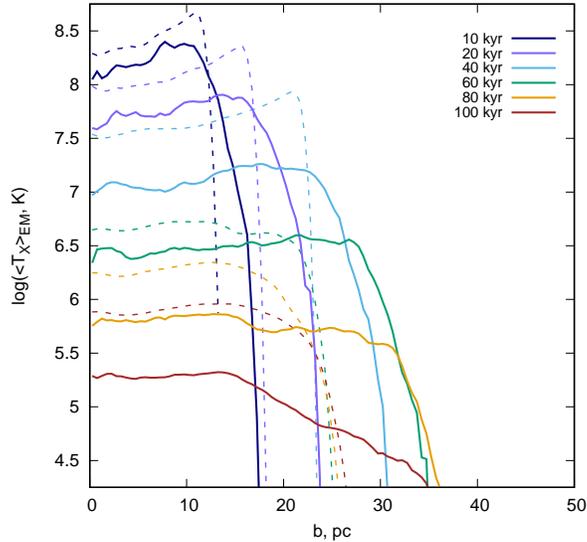}
\caption{
Gas temperature $\langle T_X(b) \rangle$, weighted with the emission measure for the SNR in an inhomogeneous  medium with $\sigma=0.2$ (dashed lines) and $\sigma=2.2$ (solid lines).
}
\label{fig-tgasx}
\end{figure*}

The observed temperature of the gas in the SNR, determined from the X-ray spectrum, depends on the conditions inside the remnant and the X-ray spectrum model used. In most cases, the single-electron nonequilibrium ionization model is applied. In this case, the temperature measured in the observations is weighted with the emission measure \citep[e.g.,][]{Leahy2019}:
\be
 \langle T_X(y,z) \rangle_{EM} = {{\sum_x T_{gas}(x,y,z) \Delta EM(x,y,z)} \over {\sum_x \Delta EM(x,y,z) }}
\ee
where $\Delta EM(x,y,z) = n_e n_p \Delta x$. 
Let us calculate the map $\langle T_X(x,y) \rangle_{EM}$ for a hot gas with $T_{gas}>3\times 10^5$K and then, averaging the map $\langle T_X(y,z) \rangle_{EM}$ over the ring layers with the impact radius $b$ according to eqn.~(\ref{eq-ave}), we obtain the profile of the average gas temperature $\langle T_X(b) \rangle_{EM} \equiv \langle T_X(b) \rangle$. Figure~\ref{fig-tgasx} shows the profile of this value for the SNR evolving in an inhomogeneous medium with dispersion $\sigma=0.2$ (dashed lines) and $\sigma=2.2$ (solid lines). It is clearly seen that the profiles of the average temperature $\langle T_X(b) \rangle$ are similar to the distributions of the surface brightness of X-ray emission of the gas in the range of $0.3-2.1$~keV (Figure~\ref{fig-surf-b}).

\begin{figure*}
\center
\includegraphics[width=8cm]{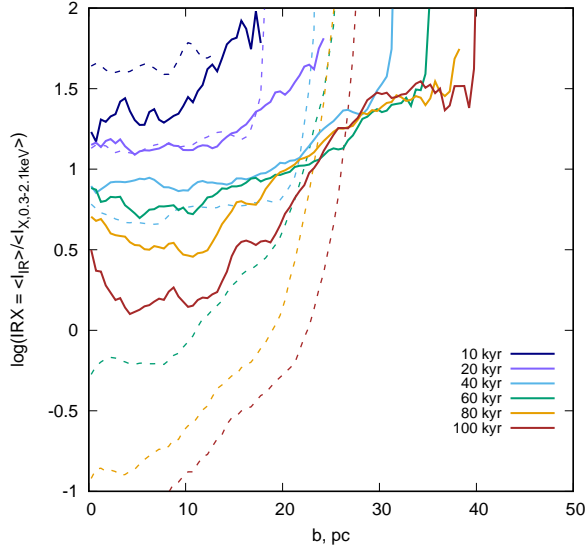}
\caption{
IRX parameter for a SNR in an inhomogeneous medium with $\sigma=0.2$ (dashed lines) and $\sigma=2.2$ (solid lines).
}
\label{fig-irx}
\end{figure*}

\subsection{$IRX$ ratio} 
\label{sec-irx}

\noindent
Using the data from Figure~\ref{fig-surf-b}, we construct the profiles of the IRX value as a function of impact parameter in Figure~\ref{fig-irx}. During the evolution of the remnant in a low inhmogeneous medium ($\sigma = 0.2$) in the adiabatic phase, IRX drops from $\sim 30$ for an age of 10~kyr to $\sim 3$ for 40~kyr at small impact distances due to a more rapid decrease of the dust emissivity owing to the slow-down in the growth of the dust mass located in the hot gas \citep[see Figure~\ref{fig-surf-b}, and][]{Dedikov2024}.  After the onset of the radiative phase, the dust predominantly passes into colder thermal phases (see Subsection~\ref{sec-evol}) and the IRX ratio drops below 1. Note that despite the effective cooling of the gas in the shell, some mass of hot gas is preserved in the central region of the SNR up to an age of $\sim 100$~kyr due to the low gas density, approximately equal to $\sim 0.1$~cm$^{-3}$. Under these conditions, the cooling time for solar metallicity gas with $T\simgt 10^6$K is  $\simgt 300$~kyr. At a larger impact distance, the ratio increases due to the growing the mass of dust along the line of sight and the decreasing the mass of hot gas. Almost vertical parts of the lines correspond to the absence of hot gas in the supernova shell at late times.

In the case of remnant evolution in a strongly inhomogeneous medium, the IRX ratio has a weak time dependence (solid lines in Figure~\ref{fig-irx}). At small impact distances, the IRX value changes by a factor of $\sim 3-4$ during evolution from 10 to 100~kyr. This is due to dust entering the hot gas from disrupted clouds and the trapping (saving) of the hot gas in the central region of the SNR (see Subsection~3.1). Starting from $\sim 40$~kyr, at distances of $b\simgt 25$~pc, the IRX ratio is approximately constant at the level of $\sim 30$. This behavior is associated with the penetration of the shock wave between the clouds, which it can no longer destroy, but is capable of heating the intercloud gas to $T$ above $10^6$K. Under such conditions, the shock wave remains adiabatic for a long period of the remnant evolution, thus, the ratio remains almost unchanged, which is similar to the behavior of the shock wave at $b\simlt 15$~pc during the first $15-20$~kyr of the remnant expansion.

\begin{figure*}
\center
\includegraphics[width=16cm]{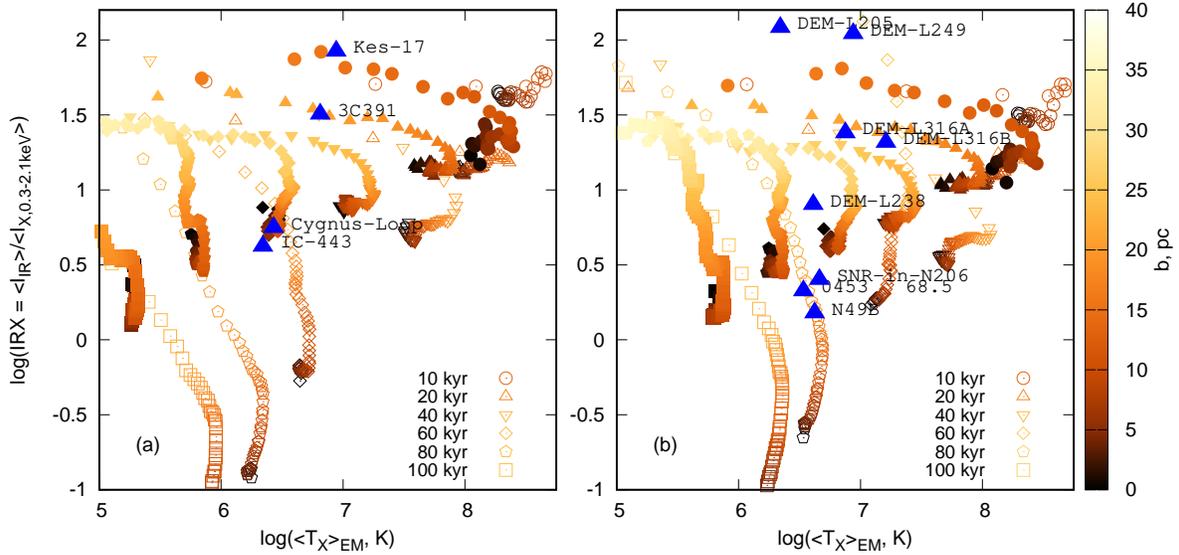}
\caption{
The diagram $T_X - {\rm IRX}$ for a SNR in an inhomogeneous medium with $\sigma=0.2$ (empty symbols) and $\sigma=2.2$ (filled symbols) for solar metallicity (panel a) and $0.5\zsun$ (panel b). Symbol types indicate the age of the SNR. The color bar encodes the impact distance. Large symbols represent the values for known SNRs in the Galaxy (a) and LMC (b) according to the data from Table~\ref{tab-sne}. The names of objects are indicated next to the corresponding symbol.
}
\label{fig-irx-tgasx}
\end{figure*}

Combining Figures~\ref{fig-tgasx} and \ref{fig-irx}, in Figure~\ref{fig-irx-tgasx}a we present the diagram $T_X - {\rm IRX}$ for the SNR evolving in the inhomogeneous medium\footnote{Unless otherwise specified, the metallicity of a gas is equal to the solar value.}. With increasing remnant age, the points shift toward lower temperatures. The loci of values for remnants evolving in environments with low and high density dispersion overlap for ages less than 20~kyr (see the upper right corner of the diagram). For older remnants, the loci of values are quite clearly distinguishable. At low dispersion, the IRX ratio drops faster. Note that some points at large impact distances (empty light symbols) correspond to high IRX values, but the IR fluxes at such distances are small.

A decrease in the metallicity of the gas leads to a longer adiabatic expansion of the remnant. Therefore, the gas temperature remain high longer, and the points on the diagram $T_X - {\rm IRX}$ shift to the high temperature region. This is clearly seen for SNRs evolving in gas with metallicity $0.5\zsun$ in Figure~\ref{fig-irx-tgasx}b. If we follow the evolution of SNRs later than 100~kyr, the values for this age shift to lower temperatures. In general, the position of the points on the diagram is determined by the onset of effective radiative losses. For higher average gas density, the points for a SNR of the same age shift to lower temperatures.

Comparing the diagrams, we can notice a slight increase in the IRX value for lower metallicity for late remnants (this is more remarkable in the model for low density dispersion). Despite the lower dust-to-gas ratio ${\cal D}$, the supernova shell remains adiabatic longer and accumulates more gas and dust.
 
\begin{table}
\caption{Supernova remnants}
\center
\begin{tabular}{lcccc}
\hline
\hline
{\bf Name}   & {\bf IRX}    & {$\bf  T_{e} (10^6 K)$}  & {\bf Age, kyr}  & {\bf Reference$^a$}\\
\hline
\multicolumn{5}{c}{SNRs in LMC}  \\
\hline
0453-68.5   & 2.12 \textpm 0.57   & 3.4   & 13		& (1,2) \\
N49B        & 1.51 \textpm 0.39   & 4.2   & 10		& (1,2) \\
DEM L205    & 121 \textpm  29.2   & 2.2   & 35		& (1,2) \\
SNR in N206 & 2.52 \textpm 0.66   & 4.6   & 25		& (1,2) \\
DEM L238    & 7.97 \textpm 2.11   & 4.1   & 10-15		& (1,2) \\
DEM L249    & 110 \textpm 30      & 8.7   & 10-15		& (1,2) \\
DEM L316B   & 20.7 \textpm 5.73   & 16.2  & $\ge $42	& (1,2) \\
DEM L316A   & 24.0 \textpm 6.55   & 7.5   & 27-39		& (1,2) \\

\hline
\multicolumn{5}{c}{SNRs in the Galaxy}  \\
\hline
3C~391       & 32		& 6.5   &  9     &  (2,3) \\
Cygnus Loop & 5.6	& 2.7   &  10    &  (2,3) \\
IC~443      & 4.2	& 2.2   &  20    &  (2,3) \\
Kes~17      & 84		& 8.8   &  8.9   &  (2,3) \\
\hline
\multicolumn{5}{l}{
$^a$ (1) \citet{Seok2013}, (2) \citet{Seok2015}, (3)  \citet{Ranasinghe2023}
} \\
\hline
\end{tabular}%
\label{tab-sne}
\end{table}

The observed values for known SNRs with ages greater than $\sim 10$~kyr are shown in the diagram $T_X - {\rm IRX}$ (Figure~\ref{fig-irx-tgasx}: large symbols in panel (a) represent objects in the Galaxy, and in panel (b) represent objects in LMC, where the gas metallicity is approximately equal to half the solar value \citep[e.g.,][]{Pei1992}. The properties of the remnants are given in Table~\ref{tab-sne}. On the one hand, it can be noted that the observed points are close to those obtained in the models. For some, in particular, for the SNR Kes~17, the position in the diagram is close to the model points corresponding to its age. On the other hand, such coincidences may be accidental, taking into account the wide scatter of points in the diagrams obtained in numerical models for a remnant of the same age depending on the impact distance, while only the average value for the entire remnant is known from observations. Also, for example, the position on the diagram for the IC~443 remnant does not correspond to its age, since it probably evolves in a remarkably denser medium, as follows from X-ray \citep[e.g.,][]{Troja2006,Troja2008} and IR \citep[e.g.,][and references therein]{Li2022IC443} observations. Note that with an increase in the average density in the medium, the points of the same age of the remnant on the diagram $T_X - {\rm IRX}$ shift to the left: during the expansion of the remnant in a denser medium, smaller $T_X$ values correspond to younger SNR. In general, when a SNR interacts with dense (molecular) clouds \cite[e.g.,][]{White1991,Chevalier1999}, the structures with complex morphology emerge in the X-ray and IR ranges \citep[see][etc.]{Braun1986,Bykov2008}, that can be apparent in higher spread of points on the diagram $T_X - {\rm IRX}$.

It is evident that in the numerical models the dispersion of the IRX value for a remnant of a given age is significant (the dependence of IRX on the impact parameter reaches several times, see Figure~\ref{fig-irx}). Therefore, a more adequate comparison with observations is possible when we get the spatial maps of the $T_X$ and ${\rm IRX}$ values for SNRs. Such maps allow us to estimate better the contribution of dust to the cooling of gas in SNRs, in comparison with the average values for the entire remnant \citep{Seok2015}.

\section{CONCLUSIONS}

\noindent
In this paper, we have considered the ratio of IR and X-ray luminosities for a supernova remnant expanding in an inhomogeneous medium -- the IRX ratio, using three-dimensional dynamics of gas and interstellar polydisperse dust particles. We have investigated the evolution of the spatial surface brightness maps of X-ray emission from hot gas inside the SNR and IR emission from the dust swept-up by the SN shell as well as the maps of average temperature of the hot gas. The results can be summarized as follows:

\begin{itemize}
 \item the IRX value changes in several times from the center to the periphery of the SNR, this is due to the dependence of cooling on dust on the radial distance;
 \item in a low inhomogeneous medium, the IRX ratio rapidly decreases during the evolution of the SNR, that can be explained by more efficient redistribution of dust in the hot gas of the SNR, while in the presence of strong inhomogeneities, dust coming from weakly destroyed fragments located in the hot gas partially compensates for the losses during thermal sputtering of grains and maintains a higher value of IR brightness in the SNRs in the radiative phase;
 \item the evolution of the locus of values for the SN remnant on the diagram $T_X - {\rm IRX}$ is determined by the onset of the radiative phase: for remnants in the adiabatic phase, the IRX values are high ($\sim 10-100$), as the remnant cools, their values shift to low values (${\rm IRX} \sim 3$), and in the case of evolution in a medium with low dispersion, lower IRX values are achieved ($\sim 0.1-1$); a decrease in the metallicity/density of the gas leads to longer adiabatic phase and higher values of both temperature $T_X$ and IRX ratio.
\end{itemize}

The IR and X-ray surface brightness distributions obtained in the numerical models are not flat depending on the impact distance (Fig.~\ref{fig-surf-b}). Such behavior can be connected with a clumpy medium. Averaging over the remnant (i.e., over the surface) leads to the loss of this information. Thus, the estimate of the contribution to gas cooling due to IR radiation of dust will be distorted. Therefore, it seems important to analyze the spatial maps of the values of $T_X$ and ${\rm IRX}$ for SNRs.

\begin{acknowledgements}
The authors are grateful to B.M. Shustov and Yu.A. Shchekinov for valuable comments and discussions, S.A. Drozdov for discussions, and I.S. Khrykin for his assistance.
\end{acknowledgements}

\bibliographystyle{aspb1}
\bibliography{p-bib1.bib}

\end{document}